\documentstyle[12pt,preprint,prl,aps,amsfonts,floats]{revtex}
\tighten 
\begin{document}  
%
\title{Berry phase for oscillating neutrinos }

\author{Massimo Blasone${}^{a,c,d}$, Peter A. Henning${}^b$  
and Giuseppe Vitiello${}^{c}$\thanks{e-mail: m.blasone@ic.ac.uk, 
P.Henning@gsi.de, vitiello@physics.unisa.it}}

\address{${}^{a}$ Blackett Laboratory, Imperial College, Prince Consort
Road, \\ London SW7 2BZ, U.K. } 
\address{${}^{b}$Institut f\"ur Kernphysik, TH Darmstadt,
         Schlo\ss gartenstra\ss e 9, \\ D-64289 Darmstadt, Germany}   
\address{${}^{c}$Dipartimento di Fisica dell'Universit\`a  
and INFN, Gruppo Collegato, Salerno \\  I-84100 Salerno, Italy} 
\address{${}^{c}$Unit\`a INFM di Salerno,  I-84100 Salerno, Italy} 

\maketitle 
\begin{abstract}  
We show the presence of a topological (Berry) phase in the time
evolution of a mixed state. For the case of mixed neutrinos,
the Berry phase is a function of the mixing angle only. 
\end{abstract}  
\vspace{0.3cm}

P.A.C.S.: 03.65.Bz, 11.10.-z, 14.60.Pq   

$$   $$

\section{Introduction}
Particle mixing and oscillations play a relevant role in high energy physics. 
In particular,
in recent years a growing interest in neutrino mixing and 
oscillations\cite{BP78}
has been developed which manifests itself both in a strong experimental
effort and in a renewed theoretical research activity. Contributions
towards the correct theoretical understanding of the field mixing
have been recently presented \cite{BV95,bosmix}.

In the present paper we show how the notion of Berry phase 
\cite{Berry} enters the physics of mixing by considering the example of
neutrino oscillations. 

Since its discovery \cite{Berry}, the Berry phase has attracted much
interest \cite{resource} at theoretical as well as at experimental
level.
This interest arises because the Berry phase reveals geometrical
features of the systems in which it appears, which go beyond the
specific dynamical aspects and as such contribute to a deeper
characterization of the physics involved.
The successful experimental findings in many different quantum systems
\cite{resource} stimulate further search in this field. 

Aimed by these motivations, we show that the geometric phase naturally
appears in the standard Pontecorvo formulation of neutrino oscillations.

Our result shows that the Berry phase associated to neutrino
oscillations is a function of the mixing angle only. 
We suggest that such a result has phenomenological relevance: since
geometrical phases are observable, the mixing angle can be (at least in
principle) measured directly, i.e. independently from dynamical
parameters as the neutrino masses and energies.

Although in the following we treat the neutrino case, we stress that
our result holds in general, also in the case of mixed bosons (Kaons,
$\eta'$s, etc..).

\section{Berry phase for oscillating neutrinos}
Let us first consider the two flavor case\cite{BP78}:
\begin{eqnarray}\nonumber
|\nu_{e}\rangle &=&
\cos\theta\;|\nu_{1}\rangle \;+\;  \sin\theta\; |\nu_{2}\rangle \;
\\ \label{nue0}
|\nu_{\mu}\rangle &=&
-\sin\theta\;|\nu_{1}\rangle \;+\;  \cos\theta\; |\nu_{2}\rangle \; .
\end{eqnarray}

The electron neutrino state at time $t$  is  \cite{BP78} 
\begin{equation}\label{nue1}
|\nu_{e}(t)\rangle \equiv e^{-i H t} |\nu_{e}(0)\rangle= 
e^{-i \omega_{1} t} \left(\cos\theta\;|\nu_{1}\rangle \;+\;
e^{-i (\omega_{2}-\omega_{1}) t}\; \sin\theta\; |\nu_{2}\rangle \;
\right),
\end{equation}
where $H |\nu_i\rangle = \omega_i |\nu_i\rangle$, $i=1,2$.
Our conclusions will also hold for the muon neutrino state, with due
changes which will be explicitly shown when necessary.
 
The state $|\nu_{e}(t)\rangle$, apart from
a phase factor, reproduces the initial state $|\nu_{e}(0)\rangle$ 
after a period $T= \frac{2\pi}{\omega_{2} - \omega_{1}}$ :
\begin{equation}\label{nue2}
|\nu_{e}(T)\rangle = e^{i \phi} |\nu_{e}(0)\rangle 
\;\;\;\;\;\;\;\;\;\;\;\;,\;\;\;\;\;\;\;\;\;\;\;\;\;
\phi= - \frac{2\pi \omega_{1}}{\omega_{2} - \omega_{1}} \,.
\end{equation}

We now show how such a time evolution does contain a purely
geometric part, i.e. the Berry phase. 
It is a straightforward calculation to separate the geometric and
dynamical phases following the standard procedure \cite{AA87}:
\begin{eqnarray}\nonumber
\beta_e&=& \phi + \int_{0}^{T}
\;\langle \nu_{e}(t)|\; i\partial_t\;|\nu_{e}(t)\rangle \,dt
\\ \label{ber1}
&=&- \frac{2\pi \omega_{1}}{\omega_{2} - \omega_{1}} +
\frac{2\pi}{\omega_{2} - \omega_{1}}(\omega_{1}\;\cos^2\theta +
\omega_{2}\;\sin^2\theta)\;= \; 2 \pi \sin^{2}\theta  \, .
\end{eqnarray} 
We thus see that there is indeed a non-zero geometrical phase $\beta$,
 related to the mixing angle
$\theta$, and that it is independent from the neutrino energies
 $\omega_1$, $\omega_2$ and masses $m_1,m_2$. 
In a similar fashion, we obtain the Berry phase for the muon neutrino
state: 
\begin{equation}\label{ber1b}
\beta_{\mu}\,= \,\phi + \int_{0}^{T}
\;\langle \nu_{\mu}(t)|\; i\partial_t\;|\nu_{\mu}(t)\rangle \,dt
\, =\, 2 \pi \cos^{2}\theta \, .
\end{equation} 
Note that $\beta_e + \beta_{\mu} = 2\pi$.
We can thus rewrite (\ref{nue2}) as
\begin{equation}\label{nue2b}
|\nu_{e}(T)\rangle = e^{i 2\pi \sin^2\theta} e^{-i \omega_{ee} T}
|\nu_{e}(0)\rangle \, ,
\end{equation}
where we have used the notation
\begin{equation}\label{omee}
\langle \nu_{e}(t)|\; i\partial_t\;|\nu_{e}(t)\rangle \,=\,
\langle \nu_{e}(t)|\; H\;|\nu_{e}(t)\rangle \,= \,
\omega_{1}\;\cos^2\theta + \omega_{2}\;\sin^2\theta
\equiv \omega_{ee}\, .
\end{equation}
We will also use
\begin{eqnarray}\label{ommm}
&&\langle \nu_{\mu}(t)|\; i\partial_t\;|\nu_{\mu}(t)\rangle \,=\,
\langle \nu_{\mu}(t)|\; H\;|\nu_{\mu}(t)\rangle \,= \,
\omega_{1}\;\sin^2\theta + \omega_{2}\;\cos^2\theta
\equiv \omega_{\mu\mu}\, ,
\\ \label{omem}
&&\langle \nu_{\mu}(t)|\; i\partial_t\;|\nu_{e}(t)\rangle \,=\,
\langle \nu_{\mu}(t)|\; H\;|\nu_{e}(t)\rangle \,= \,\frac{1}{2}
( \omega_{2}- \omega_{1})\;\sin 2\theta 
\equiv \omega_{\mu e}\, ,
\end{eqnarray}
with $\omega_{e \mu}=\omega_{\mu e}$.

In order to better understand the meaning of (\ref{ber1})-(\ref{nue2b}),
we observe that, as well known, $|\nu_{e}\rangle$ is not eigenstate of
the Hamiltonian, and
\begin{equation}\label{ovlap1}
\langle \nu_{e}(0)|\nu_{e}(t)\rangle \,=\,e^{-i \omega_{1} t}
\cos^2\theta  + e^{-i \omega_{2} t} \sin^2\theta \, .
\end{equation}
Thus, as an effect of time evolution, the state $|\nu_{e}\rangle$
``rotates'' as shown by eq.(\ref{ovlap1}). However, at $t=T$,
\begin{equation}\label{ovlap2}
\langle \nu_{e}(0)|\nu_{e}(T)\rangle \,=\,
e^{i\phi} \,=\, e^{i \beta_e} e^{-i \omega_{ee} T}\, ,
\end{equation}
i.e. $|\nu_{e}(T)\rangle$ differs from $|\nu_{e}(0)\rangle$ by a phase
$\phi$, part of which is a geometric ``tilt'' (the Berry phase) and the
other part is of dynamical origin. 
In general, for $t= T +\tau$, we have
\begin{eqnarray}\nonumber
\langle \nu_{e}(0)|\nu_{e}(t)\rangle &=&
e^{i\phi} \,  \langle \nu_{e}(0)|\nu_{e}(\tau)\rangle
\\ \label{ovlap3}
&=& e^{i 2\pi \sin^2\theta} e^{-i \omega_{ee} T}
\left( e^{-i \omega_{1} \tau} \cos^2\theta  + e^{-i \omega_{2} \tau} 
\sin^2\theta \right) \, .
\end{eqnarray}
Also notice that $\langle \nu_{\mu}(t)|\nu_{e}(t)\rangle=0$ for any
$t$. However,
\begin{equation}\label{ovlap4}
\langle \nu_{\mu}(0)|\nu_{e}(t)\rangle \,=\,\frac{1}{2}\,
e^{i\phi}  e^{-i \omega_{1} \tau}\,\sin
2\theta \left( e^{-i(\omega_2 -\omega_1) \tau} - 1\right) \qquad, \qquad
for \; t= T +\tau\, ,
\end{equation}
which is zero  only at  $t=T$. 
Eq.(\ref{ovlap4}) expresses the fact that $|\nu_{e}(t)\rangle$
``oscillates'', getting a component  of muon flavor, besides getting the
Berry phase. At $t=T$, neutrino states of different flavor are again each
other orthogonal states.

Generalization to $n-$cycles is also interesting. Eq.(\ref{ber1}) (and
(\ref{ber1b})) can be rewritten for the  $n-$cycle case as
\begin{equation}\label{ber1n}
\beta^{(n)}_{e}\,= \, \int_{0}^{nT}
\;\langle \nu_{e}(t)|\; i\partial_t -\omega_1\;|\nu_{e}(t)\rangle \,dt
\, =\, 2 \pi \,n\,\sin^{2}\theta \, ,
\end{equation} 
and eq.(\ref{ovlap3}) becomes
\begin{equation}\label{ovlapn}
\langle \nu_{e}(0)|\nu_{e}(t)\rangle \,=\,
e^{i n\phi} \, \langle \nu_{e}(0)|\nu_{e}(\tau)\rangle 
\qquad, \qquad for \; t= n T +\tau \, .
\end{equation}
Similarly eq.(\ref{ovlap4}) gets the phase $e^{i n\phi}$ instead of
$e^{i \phi}$.
Eq.(\ref{ber1n}) shows that the Berry phase acts as a ``counter'' of
neutrino oscillations, adding up $2 \pi \,\sin^{2}\theta$ to the phase
of the (electron) neutrino state after each complete oscillation.

Eq.(\ref{ber1n}) is interesting especially because it can be rewritten
as
\begin{equation}\label{ber1n2}
\beta^{(n)}_{e}\,= \, \int_{0}^{nT}
\;\langle \nu_{e}(t)|\; U^{-1}(t)\,i \partial_t \,
\Big( U(t)\;|\nu_{e}(t)\rangle\Big)
\,dt \,= \, \int_{0}^{nT}
 \langle \tilde{\nu_{e}}(t)|\;i \partial_t
|\tilde{\nu_{e}}(t)\rangle \,dt \, =\, 2 \pi \,n\,\sin^{2}\theta \, ,
\end{equation} 
with $U(t)=e^{-i f(t)}$, where $f(t)=f(0) -\omega_1 t$, and 
\begin{equation}\label{tildenue}
|\tilde{\nu_{e}}(t)\rangle \equiv U(t)|\nu_{e}(t)\rangle \, = \,
 e^{-i f(0)} \left(\cos\theta\;|\nu_{1}\rangle \;+\;
e^{-i (\omega_{2}-\omega_{1}) t}\; \sin\theta\; |\nu_{2}\rangle \;
\right)\, .
\end{equation}
Eq.(\ref{ber1n2}) actually provides
an alternative way for defining the Berry phase \cite{AA87},  which
makes use of the  state $|\tilde{\nu_{e}}(t)\rangle$ given in
eq.(\ref{tildenue}). 
In contrast with the state $|\nu_{e}(t)\rangle$, 
$|\tilde{\nu_{e}}(t)\rangle$ 
is not ``tilted'' in its time evolution:
\begin{equation}\label{ovlaptilde}
\langle \tilde{\nu_{e}}(0)|\tilde{\nu_{e}}(t)\rangle \,=\,
\langle \tilde{\nu_{e}}(0)|\tilde{\nu_{e}}(\tau)\rangle 
\qquad, \qquad for \; t= n T +\tau \, ,
\end{equation}
which is to be compared with eq.(\ref{ovlapn}).
From eq.(\ref{tildenue}) we also see that time evolution
only affects the $|\nu_2\rangle$ component of the state
$|\tilde{\nu_{e}}(t)\rangle$, so that we have
\begin{eqnarray}\nonumber
i\partial_t |\tilde{\nu_{e}}(t)\rangle &=&
(\omega_2 -\omega_1)  e^{-i f(0)} e^{-i(\omega_2 -\omega_1)t} \sin\theta
|\nu_2\rangle
\\ \nonumber
&=& (H -\omega_1) e^{-i f(0)}\left(\cos\theta\;|\nu_{1}\rangle \;+\;
e^{-i (\omega_{2}-\omega_{1}) t}\; \sin\theta\; |\nu_{2}\rangle \;
\right)
\\ \label{patilde}
&=&(H -\omega_1)|\tilde{\nu_{e}}(t)\rangle \,.
\end{eqnarray}
We thus understand that eq.(\ref{ber1n2}) directly gives us the geometric
phase because the quantity $i\langle \tilde{\nu_{e}}(t)|\dot{
\tilde{\nu_{e}}}(t)\rangle\,dt $ is the overlap of
$|\tilde{\nu_e}(t)\rangle$ with  its ``parallel transported'' at $t+dt$.

Another geometric invariant which can be considered is
\begin{equation}\label{inv1}
s\, =\, \int_0^{nT} \,\omega_{\mu e} \, dt \,= \,  \pi\,n \sin 2\theta
\,. 
\end{equation}
Since $\omega_{\mu e}$ is the energy shift from the level $\omega_{ee}$
caused by the flavor interaction term in the Hamiltonian \cite{BP78}, it
is easily seen that
\begin{equation}\label{inv1b}
\omega^2_{\mu e} = \Delta E^2 \equiv \langle\nu_{e}(t)|H^2
|\nu_{e}(t)\rangle  \, -
\, \langle\nu_{e}(t)|H |\nu_{e}(t)\rangle^2  \, ,
\end{equation} 
and then we recognize that eq.(\ref{inv1}) gives the geometric invariant
discussed in ref.\cite{AA90}, where it is defined quite generally as
$s=\int \Delta E(t) dt$. It has the advantage to be well defined
also for systems with non-cyclic evolution.

We now consider the case of three flavor mixing. Consider again the
electron  neutrino state at time $t$ \cite{BV95}:
\begin{eqnarray}\nonumber
|\nu_{e}(t)\rangle &= &
e^{-i \omega_{1} t} \left(\cos\theta_{12}\cos\theta_{13}
\;|\nu_{1}\rangle \;+\;
e^{-i (\omega_{2}-\omega_{1}) t}\; \sin\theta_{12}\cos\theta_{13}
\; |\nu_{2}\rangle \; \;+\; \right. 
\\   \label{nue3}
&& \left. \qquad \qquad   \qquad \qquad  \qquad \qquad 
 \qquad \qquad  \qquad \qquad e^{-i (\omega_{3}-\omega_{1}) t} 
e^{i\delta}\; \sin\theta_{13} \; |\nu_{3}\rangle \;  \right) \, ,
\end{eqnarray}
where $\delta$ is the analogous of the CP violating phase of the CKM
matrix. 
Let us consider the particular case in which the two frequency
differences are proportional: $\omega_{3}-\omega_{1} = q
(\omega_{2}-\omega_{1})$, with $q$ a rational number. In this case the
state (\ref{nue3}) is periodic over a period  $T= \frac{2\pi}{\omega_{2}
- \omega_{1}}$ and we can use the previous definition of Berry phase:
\begin{equation}\label{3flav1}
\beta\,= \,\phi + \int_{0}^{T}
\;\langle \nu_{e}(t)|\;H\;|\nu_{e}(t)\rangle \,dt
\;= \; 2 \pi \left( \sin^{2}\theta_{12}\cos^2\theta_{13}  + q 
\sin^{2}\theta_{13} \right)   \, ,
\end{equation}
which of course reduces to the result (\ref{ber1}) for
$\theta_{13}=0$. 
Eq.(\ref{3flav1}), however, shows that $\beta$ is not completely free
from dynamical parameters since the appearance in it of the parameter
$q$. 

Although because of this, $\beta$ is not purely geometric, nevertheless
it is interesting that it does not depend on the specific frequencies
$\omega_i, \; i=1,2,3\,,$ but on the ratio of their differences only. 
This means that we have now (geometric) classes labelled by $q$.

It is  in our plan to calculate the geometric invariant $s$
for the three flavor neutrino state: this requires
consideration of the projective Hilbert space in the line of
ref.\cite{AA90,dorje}. 

\section{Final remarks and conclusions}

The geometric phase is generally associated with a parametric dependence
of the time evolution generator. In such cases, the theory exhibits a
gauge-like structure which may become manifest and characterizing for
the physical system, e.g. in the Bohm-Aharonov effect\cite{baeff}.

It is then natural to ask the question about a possible gauge structure
in the case considered in this paper. Let us see how, indeed, a
covariant derivative may be here introduced.

Let us consider the evolution of the mass eigenstates
\begin{equation}\label{ga1}
i \partial_t |\nu_i(t)\rangle \, =\, H
\,|\nu_i(t)\rangle   \, ,
\end{equation}
where $i=1,2$. 
These equations are  invariant under the following (local in time)
gauge transformation 
\begin{equation}\label{ga2}
|\nu_i(t)\rangle \, \rightarrow \, |\tilde{\nu_i}(t)\rangle \equiv
U(t)|\nu_i(t)\rangle\, =\,   e^{-if (t)} |\nu_i(t)\rangle   \, ,
\end{equation}
provided $[H,U(t)]=i \partial_t U(t)$, i.e.
\begin{equation}\label{ga5}
U^{-1}(t)H U(t) \, =\,
H \, + \,  U^{-1}(t)i \partial_t U(t) 
=\,H \, + \, \partial_t f(t)
\, .
\end{equation}

This suggests that, by rewriting (\ref{ga1}) as
\begin{equation}\label{ga1b}
\left( i \partial_t \, - \, H\right)|\nu_i(t)\rangle \, =\, 0
 \, ,
\end{equation}
we can consider $D_t\equiv \partial_t \, +  \, i H $ as the ``covariant
derivative'':
\begin{equation}\label{ga3}
D_t \, \rightarrow \, D'_t \,
=\, U(t) D_t U^{-1}(t) \,.
\end{equation}
We have indeed
\begin{equation}\label{ga6}
i  D'_t |\tilde{\nu_i}(t)\rangle 
\, =\, i  D'_t  U(t) |\nu_i(t)\rangle \, =
\, i  U(t) D_t |\nu_i(t)\rangle \, = \, 0 \, ,
\end{equation}
which in fact expresses the invariance of eq.(\ref{ga1}) under
(\ref{ga2}). 

We thus see that the time dependent canonical transformation of the
Hamiltonian eq.(\ref{ga5}) and eq.(\ref{ga2}) play the
role of a local (in time) gauge transformation. Note that the state
$|{\tilde \nu}_e(t)\rangle$ of eq.(\ref{tildenue}) 
is a superposition of the
states $|{\tilde \nu}_i(t)\rangle$.

The role of the ``diabatic'' force arising from the term $U^{-1}(t)i
\partial_t U(t)$ has been considered in detail elsewhere \cite{SVW98}.

Summarizing, we have shown that there is a Berry phase built in  in the
neutrino oscillations, we have explicitly computed it in the cyclic
two-flavour case and in a particular case of three flavor mixing.
The result also applies to other (similar) cases of 
particle oscillations.

We have noticed that a measurement of this Berry phase would give a
direct measurement of the mixing angle independently from the values of
the masses.

The above analysis in terms of ``tilting'' of the state in its time
evolution, parallel transport and covariant derivative also suggests
that field mixing may be seen as the result of a curvature in the state
space. The Berry phase appears to be a manifestation of such a
curvature. 

Finally, we remark that the recognition of the geometric phase
associated to mixed states also suggests to us that a similar geometric
phase also occurs in entangled quantum states which can reveal to be
relevant in completely different contexts than particle oscillations,
namely in quantum computation \cite{qcomp}.

\smallskip
\section{Acknowledgements}
We thank M.Nowak and D.Brody for fruitful discussions.
M.B. and G.V. acknowledge INFN, INFM, MURST and ESF for support.
 

\end{document}